\documentclass[submit,article,moreauthors]{mdpi}
\usepackage[T1]{fontenc}
\usepackage{lmodern}
\usepackage[utf8]{inputenc}

\usepackage{array}
\usepackage{tabularx}
\usepackage{booktabs}
\usepackage{adjustbox}

\AtBeginDocument{%
  \setlength{\tabcolsep}{4.5pt}%
  \renewcommand{\arraystretch}{1.1}%
  \newcolumntype{P}[1]{>{\raggedright\arraybackslash\hspace{0pt}}p{#1}}
  \newcolumntype{Y}{>{\raggedright\arraybackslash\hspace{0pt}}X}
}

\firstpage{1} 
\pubvolume{1}
\issuenum{1}
\articlenumber{0}
\pubyear{2025}
\copyrightyear{2025}

\datereceived{ } 
\daterevised{ } 
\dateaccepted{ } 
\datepublished{ } 

\hreflink{https://doi.org/} 

\Title{A Systematic Review: Affective Perception on Urban Facade}
\TitleCitation{Title}

\Author{Chenxi Wang $^{1}$, Haining Ding $^{2}$ and Michal Gath-Morad $^{2,}$*}
\AuthorNames{Chenxi Wang, Haining Ding and Michal Gath-Morad}

\isAPAStyle{%
  \AuthorCitation{Wang, C.; Ding, H.; Gath\textendash Morad, M.}
}{%
  \isChicagoStyle{%
    \AuthorCitation{Wang, Chenxi, Haining Ding, and Michal Gath\textendash Morad.}
  }{
    \AuthorCitation{Wang, C.; Ding, H.; Gath\textendash Morad, M.}
  }
}

\address{%
$^{1}$ \quad Cambridge Cognitive Architecture, Department of Architecture, University of Cambridge, UK\\
$^{2}$ \quad NeuroCivitas Lab for NeuroArchitecture, Centre for Research in the Arts, Social Sciences and Humanities (CRASSH), University of Cambridge, UK}

\abstract{Architectural facades critically shape affective perception in urban environments. Here, affect is understood as a multidimensional psychological construct encompassing valence (pleasure–displeasure) and arousal (activation–deactivation). Despite growing interest in affective responses to the built environment, the affective impact of urban architectural facades remains under-theorized. This study conducts a systematic review of 61 works, guided by the PRISMA framework, to identify which facade attributes most strongly predict affective responses operationalized as valence and arousal. Through multi-scalar synthesis and knowledge mapping, the review highlights complexity, materiality, symmetry, and bibliophilic integration as consistent predictors of affective perception across urban, building, and detail levels. Computational tools such as eye-tracking, CNN-based analysis, and parametric modeling are increasingly employed, yet remain fragmented and often overlook intangible dimensions like narrative coherence and cultural symbolism. By consolidating cross-disciplinary evidence, this review proposes a theoretical model linking physical design features to affective outcomes, and identifies methodological gaps, particularly the lack of integrative, mixed-method approaches. The findings offer a foundation for affect-aware facade design, advancing evidence-based strategies to support psychological well-being in urban contexts.
}


\usepackage{booktabs}   
\usepackage{tabularx}   
\usepackage{caption}
\renewcommand{\arraystretch}{1.1}

\begin{document}

\section{Introduction}

Recent policies and research in the United Kingdom reflect a heightened emphasis on the interplay between architectural appeal aesthetic values, and well being. The National Design Guide (2021) \citep{MHCLG2021} underscores the critical role of materials, forms, and environmental factors in fostering psychological comfort through architectural design. The \textit{Building Better, Building Beautiful Commission} \citep{BBBBC2020} further explores how aesthetic considerations in design can improve urban happiness and public health. Research shows that attractive urban spaces, particularly those with well-designed architecture, can evoke pleasure, reduce stress, and foster positive affective states \citep{Lindal2013, HollanderAnderson2020}. A 2023 Thinks Insight survey found that 76 \% of UK respondents believe the appearance of buildings affects their mental health, with 75\% reporting positive moods when engaging with visually appealing facades \citep{Thinks2023}. On the other hand, visual disorder and monotony in urban settings tend to amplify anxiety, induce discomfort, and compromise overall well-being \citep{Kotabe2016, Parsons1991, Lindal2013}. Here, affective states, defined as through two main axes, valence (pleasure level) and arousal (activation intensity), have been widely established as fundamental dimensions of affective states in environmental psychology and affective neuroscience \citep{Russell1999, Posner2005}.

From an evolutionary psychology perspective, Sussman and Hollander \citep{Sussman2015} identify four characteristics underlying biophilic preference in human affective perception: Edges - articulation of the facade along paths and street corridors; Shapes - bilateral symmetry akin to facial symmetry; Patterns - repetition of motifs fostering coherence; and Narrative - historical significance cultivating a sense of place. Among urban visual elements, façades are pivotal as the primary interface between individuals and the built environment \citep{Lindal2013, HollanderAnderson2020}. Facades not only regulate environmental performance (e.g., daylight and thermal dynamics; \citep{Yi2019}), but also act as cognitive stimuli, evoking affective evaluations through their geometric, textural, and material qualities \citep{KaplanKaplan1982, Sussman2015}. For example, façades with high permeability, diverse materials, and rhythmic patterns evoke pleasure and engagement, while transitional features such as balconies or awnings enhance emotional comfort \citep{HollanderAnderson2020}. Similarly, textural complexity and dynamic shapes stimulate aesthetic arousal \citep{KaplanKaplan1982, Nasar1994}, whereas monotonous façades can induce stress and disengagement \citep{Ellard2012}.

Although the relationship between the built environment and affective states has been extensively documented \citep{Parsons1991, Kaplan1995, Tost2019, Zhang2021, Kaklauskas2021, White2019, Valentine2025}, the specific role of architectural facades in shaping human affective perception remains not adequately addressed \citep{HollanderAnderson2020}, the specific role of architectural facades in shaping human affective perception remains not adequately addressed \citep{HollanderAnderson2020}. Existing studies employing surveys and interviews, although valuable, often rely on subjective assessments that face challenges of scalability and cross-regional comparability \citep{Nasar1994, Dunstan2005}. To address these limitations, recent advances in urban informatics and computational design technologies offer promising alternatives. Tools such as Street View Imagery (SVI) allow scalable and systematic analysis of urban facades, enabling researchers to capture visual characteristics and assess their affective impact \citep{Biljecki2021, Zhang2018}.

\section{Research Question}
Hence, this essay aims to broaden our understanding of the human perception of building exteriors. The purpose of this paper is to address the following research questions. How do architectural facades influence affective perception in urban environments?

To elaborate, which facade attributes, across streets, buildings, and detail levels, most significantly influence affective perception, particularly in terms of arousal and valence, in urban spaces? In addition, how can interdisciplinary methodologies, including psychological frameworks, computer vision tools, and empirical studies, advance the understanding of the relationship between affective perception and urban facades?

\section{Theoretical Model of Affect}

\begin{figure}[H]
\includegraphics[width=14.0 cm]{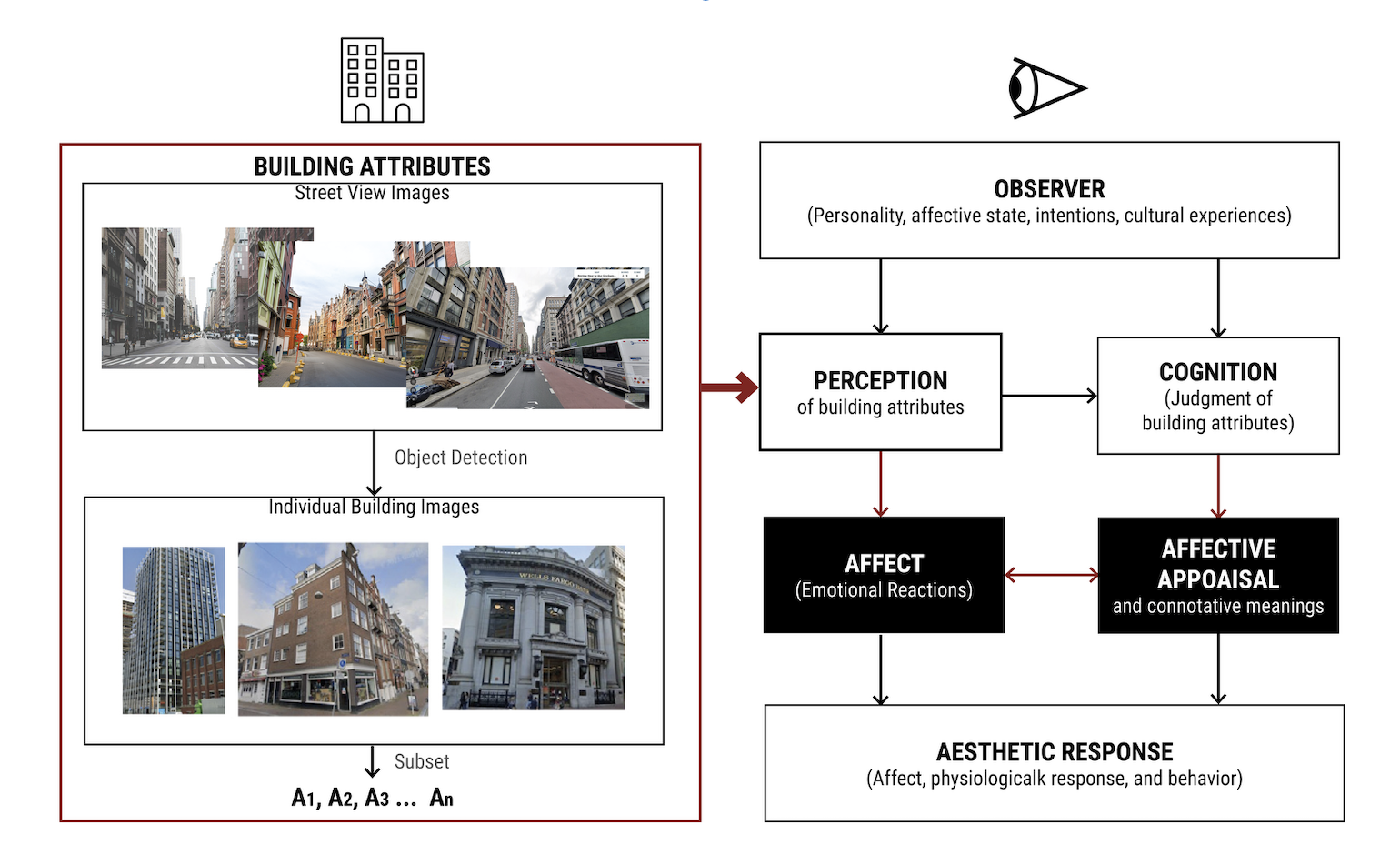}
\caption{Conceptual framework linking building attributes, perception, and aesthetic responses. Source: Author.\label{fig1}}
\end{figure}

This literature review investigates how architectural facades evoke affective responses, processed through perception, cognition, and affective appraisal. The focus here is on affect and affective perception, emphasizing initial emotional reception closely related to visual stimuli. Affect, illustrated in Figure~\ref{fig1}, constitutes a central feature of human experience \citep{Barrett2009}, and is widely modeled along two dimensions: valence (ranging from pleasant to unpleasant) and arousal (ranging from high to low activation, such as alert versus tired) \citep{Russell1999}. As an immediate and often unconscious emotional response to stimuli, affect functions as a neurophysiological barometer of one’s relationship with the environment \citep{Scherer2005, Posner2005}. For example, vibrant and textured facades may evoke excitement and high arousal, whereas monotonous or visually cluttered designs tend to induce discomfort or low valence \citep{Nasar1994, Ellard2012, Zhang2021}. Moreover, as Pessoa \citet{Pessoa2008emotion, Pessoa2013brain} argues, affect and cognition are deeply integrated, operating synergistically across perceptual and evaluative domains. While cognition plays a critical role, its complexity exceeds the scope of this paper and will not be extensively discussed here. The proposed theoretical framework (Figure~\ref{fig1}) outlines the psychological mechanisms of facade perception in urban settings.

\unskip
\vspace{1em} %

The affective response process begins with perception, where sensory modalities such as vision and touch register physical facade attributes such as color, texture, and geometry \citep{Gibson1979, Goldstein2014}. These sensory impressions and perception feed as raw data into cognitive processes, which integrate memory, expectations, and contextual knowledge to enable higher-order evaluations and judgment \citep{neisser1967cognitive, Anderson2005}. However, affective responses often occur before and independently of cognition, highlighting their immediacy and automaticity \citep{Zajonc1984}. For instance, rapid emotional reactions to visual cues like symmetry or rhythmic patterns can arise without conscious deliberation \citep{Zajonc1984}. While affect is primarily automatic, affective appraisal adds a cognitive layer of evaluation. This process integrates initial emotional reactions with cognitive elements such as contextual and experiential factors, allowing individuals to interpret and assess their environment in a more nuanced way \citep{Lazarus1991, Scherer2005}. For example, a facade perceived as 'inviting' or 'oppressive' is influenced not only by sensory impressions but also by deeper cognitive evaluations shaped by cultural norms, past experiences, and expectations. These evaluations guide behavioral decisions, such as whether an environment feels safe, engaging, or aesthetically pleasing, thereby influencing how individuals behave with urban spaces.

\begin{figure}[H]
\includegraphics[width=14.0 cm]{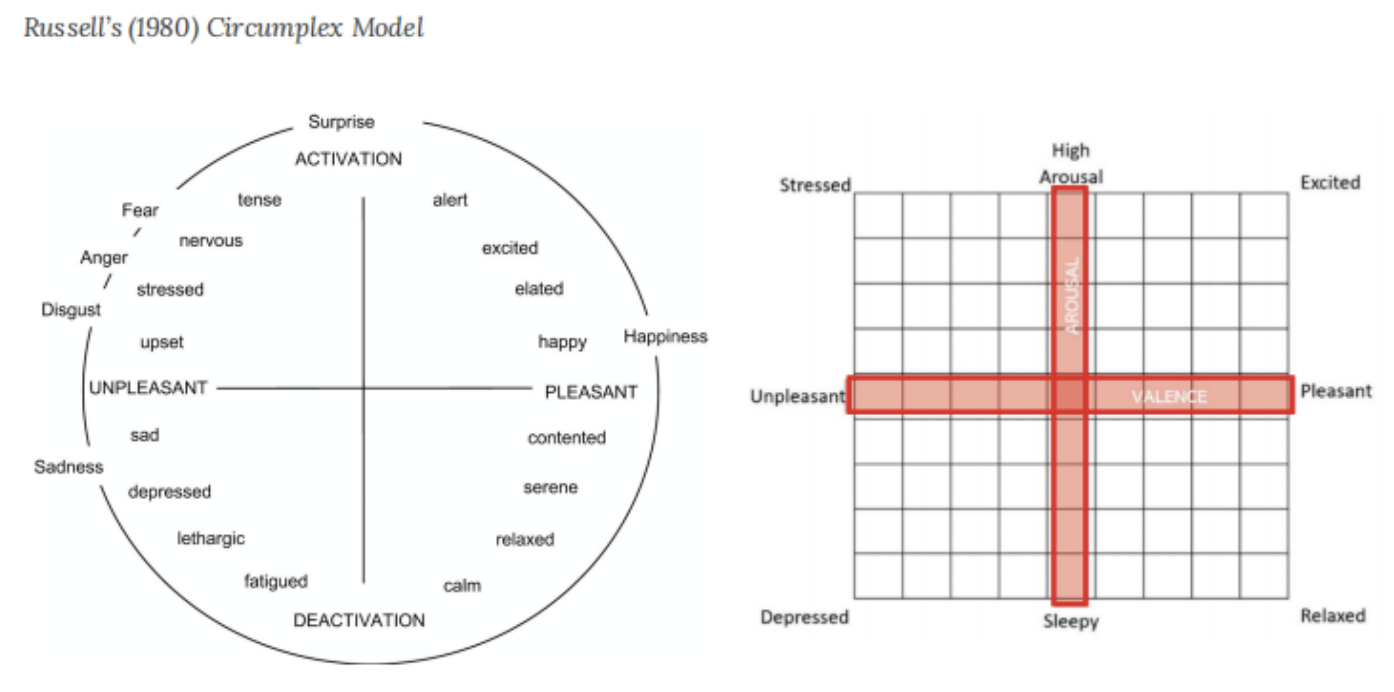}
\caption{Russell’s (1980) Circumplex \citep{Russell1989} illustrating affect dimensions of valence and arousal, and a modified affect grid used for study purposes.
\label{fig2}}
\end{figure}   
\unskip
\vspace{1em} %

Note that affective perception is not universal. Individual differences, such as anxiety sensitivity or agoraphobic tendencies, strongly shape the perception of urban spaces. Anxiety-prone individuals often exhibit heightened arousal in everyday environments, while those prone to agoraphobia experience discomfort in open or crowded settings \citep{Lederbogen2011, Evans2003}. More broadly, neuroimaging evidence shows that individual variation critically moderates the affective benefits derived from urban green space exposure \citep{Tost2015}. Acknowledging these differences is crucial for developing a nuanced understanding of the perceptual dimensions of urban environments, highlighting the urgency for cities to account for diverse affective responses.

\section{Methods}

The basis for the methodology adopted in this study is the Preferred Reporting Items for Systematic Reviews and Meta-Analyses (PRISMA), a widely adopted structured approach in evidence-based research aimed at increasing transparency, reproducibility, and rigor in review processes. The PRISMA flow diagram outlines four key stages: Identification, Selection, Eligibility, and Inclusion,to transparently report the selection process of studies in systematic reviews. Keywords related to facade, affective perception, and urban context were strategically selected for searches across major academic databases. After duplication, screening of the title and abstract and full text screening, 61 studies were selected that met the predefined inclusion criteria.

The analysis also proceeded in two phases: 1) descriptive analysis: This phase identified, categorized, and organized the selected works into thematic areas of focus, namely facade attributes and their associated affective impacts; and 2) explanatory synthesis: This phase connected isolated research points by describing the relationships between influencing variables and establishing commonalities and differences.

\subsection{Search Strategy} 

\begin{figure}[H]
\begin{adjustwidth}{-\extralength}{0cm} 
    \centering
    \includegraphics[width=\fulllength]{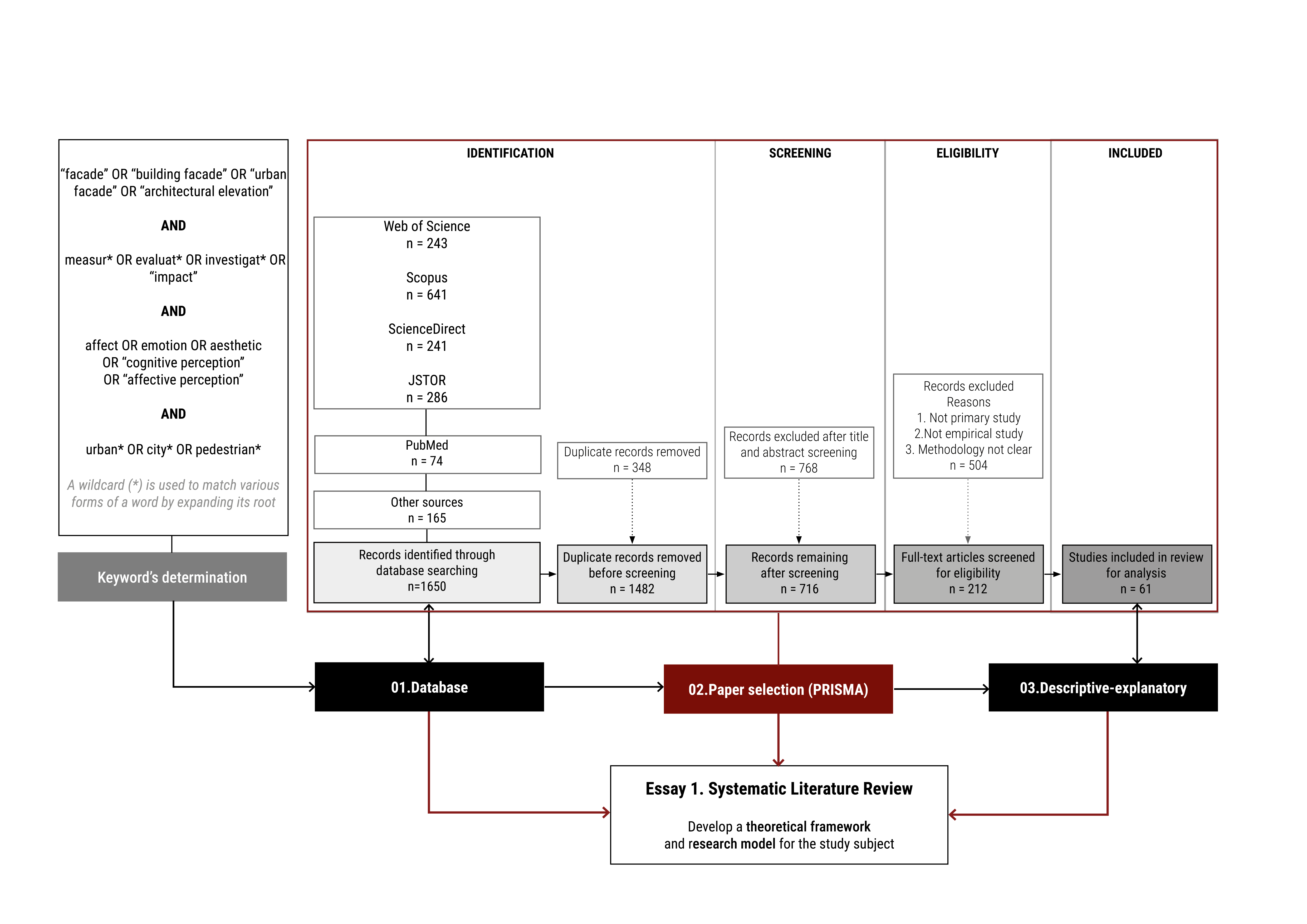}
    \caption{PRISMA flow diagram for systematic literature review, outlining database searches, screening, and paper selection process for the study. Source: Author.\label{fig3}}
\end{adjustwidth}
\end{figure}

The systematic literature review (Figure~\ref{fig3}) follows a structured process for the determination of keywords, the search of databases, and the subsequent selection stages using the PRISMA framework. The process begins with the determination of keywords that are designed to capture a broad yet precise range of literature relevant to the research objectives. These include terms such as 'facade', 'building facade', 'urban facade' and 'building exterior', alongside others such as 'measure', 'evaluate', 'impact', 'cognitive perception', 'affective perception', and 'aesthetic.' Contextual terms such as 'urban', 'city', and 'pedestrian' are included to ensure that the focus remains on urban environments and their interaction with architectural facades. A wildcard (*) is utilized to broaden the search scope by incorporating all variations of root words, maximizing the inclusion of the search results.

The search was conducted across below major databases: Web of Science, Scopus, ScienceDirect, JSTOR, PubMed, and other sources, yielding a total of 1,650 records. During the identification phase, duplicate entries (n=348) were removed, leaving 1,482 unique records for further evaluation. The screening phase involved a detailed review of titles and abstracts to exclude studies that did not align with the research focus, such as those lacking relevance, empirical basis, or a clear methodological framework. This step resulted in the exclusion of 768 records, leaving 716 articles for full-text assessment.

The eligibility stage applied stringent inclusion and exclusion criteria (Table.01) to the 716 remaining articles. Studies were included only if they were primary and empirical research with clearly defined methodologies, directly addressing the study’s thematic focus. Articles that were not empirical, lacked primary data or had unclear methods were excluded (n=504), leaving 212 studies for detailed eligibility evaluation. Of these, 61 studies were deliberately selected as suitable for inclusion in the systematic literature review.

While the PRISMA framework enhances methodological transparency and systematic rigor, its application outside its intended domain necessitates critical reflection to ensure that the modifications made align with the research objectives and do not compromise validity. Additionally, the responsibility of adapting and applying PRISMA solely falls on one author (me), further underscoring the importance of cautious and thoughtful application to mitigate any potential methodological constraints or biases. However, this systematic approach, with these considerations in mind, ensures the relevance, methodological soundness, and quality of the selected studies, offering a strong basis for developing insights and advancing knowledge in the field.

\subsection{Synthesis and Analysis} 
Employing the PRISMA framework ensures a systematic selection of high-quality literature, that providing a robust foundation for subsequent analyses. CiteSpace also facilitates the visualization of research trends and knowledge structures within a specific domain. By extracting high-frequency terms from titles, abstracts, and keywords, CiteSpace constructs a co-occurrence matrix to quantify the frequency with which terms appear together. This matrix is then visualized as a network map, where nodes represent keywords, with their size proportional to term frequency, and edges indicate cooccurrence relationships, with thickness corresponding to cooccurrence frequency. CiteSpace’s unique strength lies in its ability to integrate co-occurrence analysis with temporal mapping, allowing researchers to visualize the evolution of knowledge structures and identify shifting paradigms and emerging frontiers with precision.

\begin{figure}[H]
\includegraphics[width=12.0 cm]{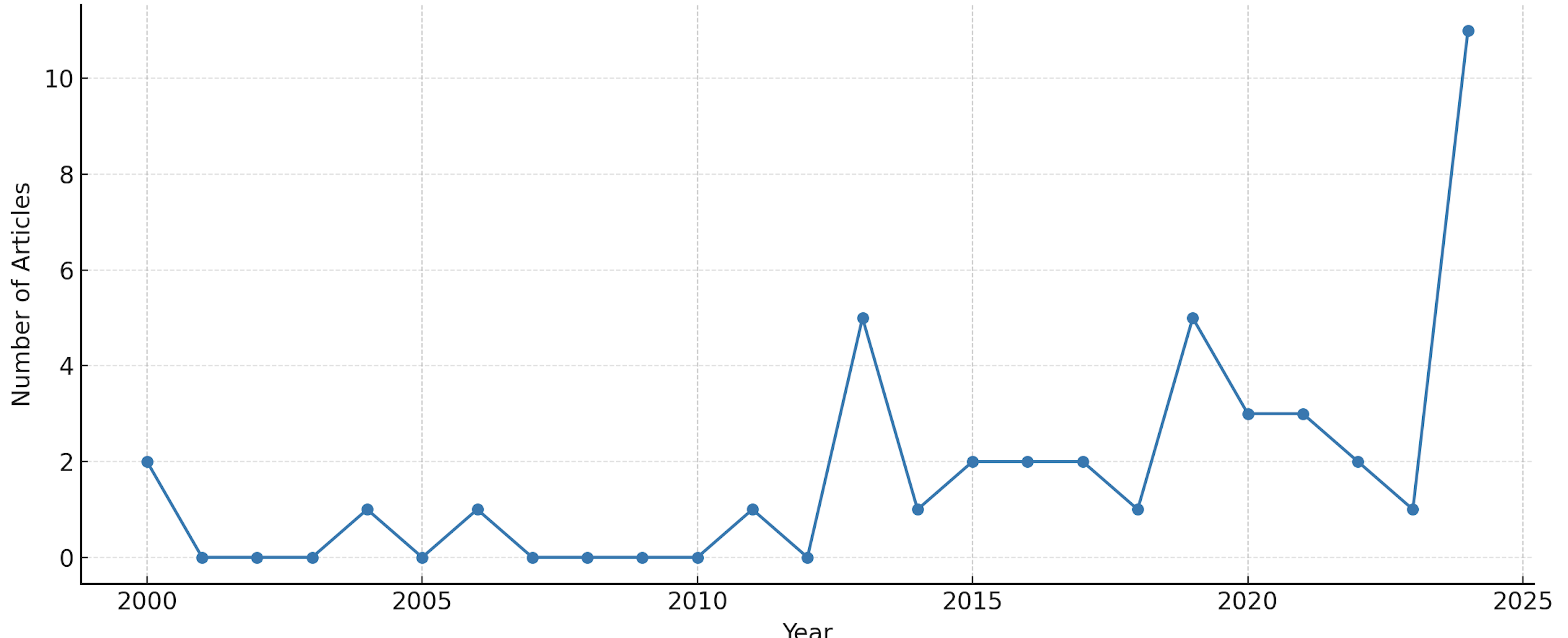}
\caption{The number of publication each year 2000 -2025 (include in this paper). Source: Author.\label{fig4}}
\end{figure}   
\unskip
\vspace{1em} %

\subsubsection{Evolution and Trends in Facades Perception} 

\begin{figure}[H]
\begin{adjustwidth}{-\extralength}{0cm} 
    \centering
    \includegraphics[width=\fulllength]{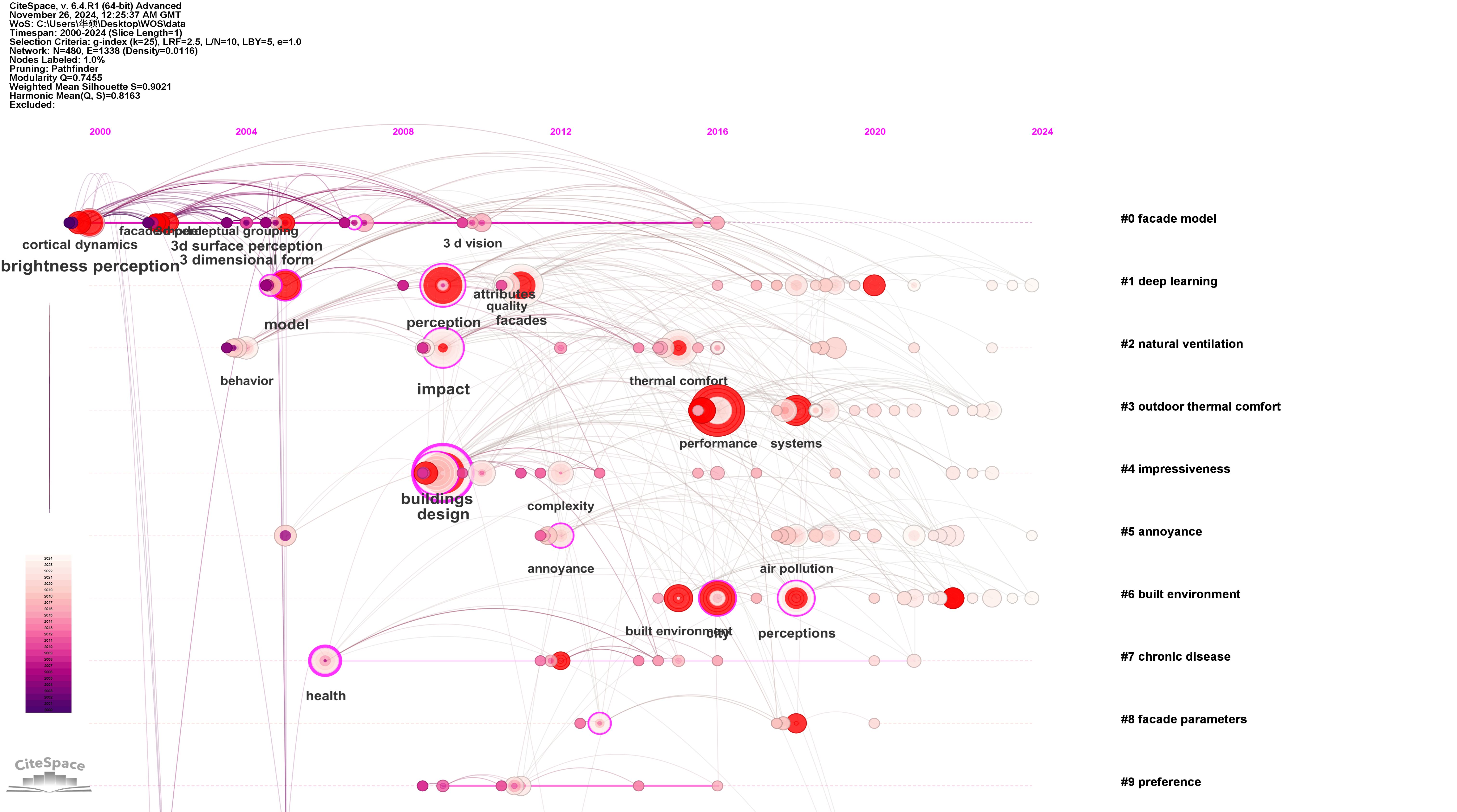}
    \caption{Knowledge mapping of facade-related research from 2000 to 2024 using CiteSpace. The map visualizes research topics, citation relationships, and clusters. Source: Author.\label{fig5}}
\end{adjustwidth}
\end{figure}

Figure,~\ref{fig5} collectively present a cocitation analysis of architectural facade research, mapping the progression of topics and their interconnections from 2000 to 2024. Figure~\ref{fig4}  provides an overview of the chronological development of research topics related to architectural facades and their impact on human perception, spanning the years 2000 to 2024. The horizontal axis represents time, showcasing the chronological progression and development of various research topics. The nodes in the network represent specific research topics or keywords, with the size of each node indicating its frequency of citation. Larger nodes signify topics that are highly cited and represent key focal areas within the field. The color of the nodes provides temporal context, with red indicating more recent studies, while yellow represents older research. Nodes with pink or purple borders are identified as having high between-ness centrality, suggesting that these topics serve as critical bridges connecting different subfield within the domain.

\subsubsection{Co-Citation Network of Facades Perception} 

The timeline in Figure~\ref{fig4}  traces the evolution of facade research, beginning with foundational studies on “3D surface perception,” “brightness perception,” and “cortical dynamics” in the early 2000s. These efforts established a framework for understanding how materiality, texture, and dimensions shape perception, forming the basis of Cluster \#0: Facade Model Figure~\ref{fig5}. Subsequent clusters, such as \#8: Facade Parameters and \#9: Preference, extend this understanding by integrating user-centered approaches to balance technical precision with aesthetic evaluation. The adoption of Cluster \#1: Deep Learning reflects a shift to data-driven methods, leveraging artificial intelligence to optimize facade design, aligning computational precision with user preferences.

\begin{figure}[H]
\begin{adjustwidth}{-\extralength}{0cm} 
    \centering
    \includegraphics[width=\fulllength]{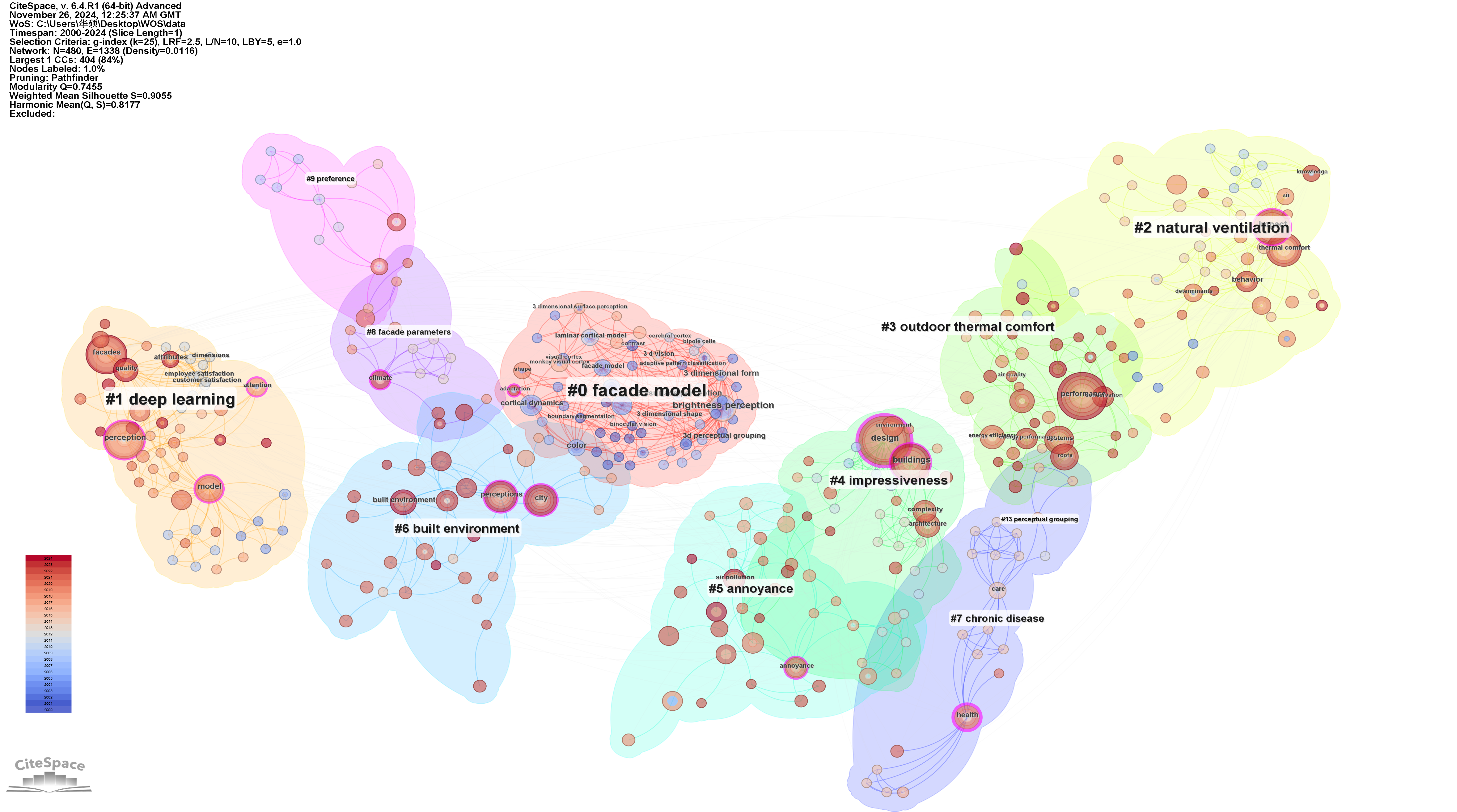}
    \caption{Clustered knowledge mapping of facade-related research topics from 2000 to 2024 using CiteSpace. Source: Author.\label{fig6}}
\end{adjustwidth}
\end{figure}

The cocitation network highlights thematic connections across clusters, with cluster \#4: Impressiveness and Cluster \#5: Annoyance that reveals the cognitive and emotional impacts of the facades. Positive responses such as admiration are explored in cluster \#4, while cluster \#5 examines stress and discomfort caused by poor design. Broader themes in Cluster \#6: Built Environment and \#7: Chronic diseases link facade design to urban systems, environmental psychology, and public health, demonstrating their dual function as aesthetic elements and urban mediators. Together, Figs. 04 and 05 capture a progression from morphological studies to computational innovation and interdisciplinary integration, underscoring the multifaceted role of facades in urban sustainability and human experience. Building on the cocitation and timeline analysis derived from the PRISMA-selected literature, the subsequent descriptive-interpretive elaborations will critically refine and expand these foundations with enhanced precision and depth. Comprehensive analyses and detailed research processes are provided in the Appendix.

\begin{table}[H]
\tablesize{\scriptsize}
\caption{Research fields and representative journals on affective perception of façades.}
\label{tab:fields}
\begingroup\hyphenpenalty=10000\exhyphenpenalty=10000\sloppy
\begin{adjustwidth}{-\extralength}{0cm}
\begin{tabularx}{\fulllength}{L L L X X X}
\toprule
\textbf{Research Field} & \textbf{Paper Count} & \textbf{Representative Journals} & \textbf{Reference} \\
\midrule
1. Environmental Psychology \& Cognitive Studies & 8 & \textit{Environment and Behavior, Journal of Environmental Psychology, Psychology of Aesthetics, Creativity, and the Arts} & \citep{Gifford2000, Heath2000, Reber2004, White2011, Lindal2013, Zhang2021, Kaklauskas2021, Grzywacz2024} \\

2. Architectural Aesthetics \& Design & 7 & \textit{Architectural Science Review, Architectural Engineering and Design Management, Indoor and Built Environment, Buildings} & \citep{Shemesh2017, Aydin2022, Piroozfar2015,Ghomeshi2013,Wang2024b, montanana2024prestige, Zhu2024} \\

3. Urban Design \& Planning & 5 &  Landscape and Urban Planning, Journal of Urban Design, Journal of Urbanism, Archnet-IJAR & \citep{OConnor2006, Hollander2019, Chung2022, Hollander2020, AbuQadourah2024}\\

4. Historical \& Cultural Architectural Studies & 5 & \textit{Frontiers of Architectural Research, Land, Procedia Engineering} & \citep{Farida2013, Ariannia2024,Naceur2013,DeLaColina2016,Kaklauskas2021} \\

5. Technology-Driven Architectural \& Urban Analysis & 4 & \textit{ACM Transactions on Graphics, Building and Environment, International Journal of Architectural Computing} & \citep{Vangorp2013,Yi2019,Liang2024,Ghozatlou2024}\\

6. Acoustics \& Visual Perception & 4 & \textit{Building Acoustics, Journal of Eye Movement Research, Fresenius Environmental Bulletin } & \citep{Calleri2018, Calleri2017, Beder2024, Asur2020} \\
\addlinespace

\bottomrule
\end{tabularx}
\end{adjustwidth}
\endgroup
\end{table}

\section{Findings}
\subsection{Traditional approaches in architecture facade evaluation} 

In recent decades, the study of sensory interactions between observers and architectural facades, grounded in environmental psychology, has gained prominence in built environment research \citep{Parsons1991, Craik1973, Zhang2021, Kaklauskas2021, Liang2024, Valentine2025}. Facade design, including its form, texture, and visual complexity, has been recognized for its role in shaping both affective perception and aesthetic appeal \citep{Nasar1994, Kaplan1995, Appleyard1972, Daniel2001}. Empirical studies have further demonstrated how specific facade attributes influence affect and well-being \citep{Nasar1994, Lindal2013, Dunstan2005}. For example, Lindal and Hartig \citet{Lindal2013} asked participants to evaluate architectural scenes varying in roofline silhouette, surface ornamentation, and building height, showing systematic differences in restorative potential.

For the purposes of operationalization assessments of ‘high-quality’ facades, traditional approaches have isolated the attributes driving affective responses, commonly distinguished as tangible (physical) and intangible (conceptual) attributes. Of these, coherence, complexity, enclosure, and mystery have been central to traditional assessments of aesthetic and emotional responses to architectural exteriors. These properties provide a structural context through which to analyze the way facade design affects human affective experience by connecting physical characteristics with emotional evaluations. For example, mystery represents the quality of a facade that provides implicit information or visual cues to stimulate the observer’s curiosity and desire to explore \citep{Herzog1992}. Meanwhile, coherence refers to the extent of visual integrity and understandability and facilitates better processing and memory of architectural forms \citep{Gifford2000, Nasar1994}. Whereas moderate complexity offers enhanced visual participation and emotional appeal, excessive complexity tends to overwhelm observers, thereby reducing aesthetic pleasantness \citep{Rapoport1990, Heath2000}. More specifically, perceived complexity often shows a U-shaped relationship with preference, where moderate complexity is the most favored \citep{Imamoglu2000, Akalin2009}. The enclosure is related to restoration through the feeling of physical or visual confinement. As Lindal and Hartig \citet{Lindal2013} note, a strong enclosure combined with medium entropy made the restorative experience high. 

\begin{figure}[H]
\begin{adjustwidth}{-\extralength}{0cm} 
    \centering
    \includegraphics[width=\fulllength]{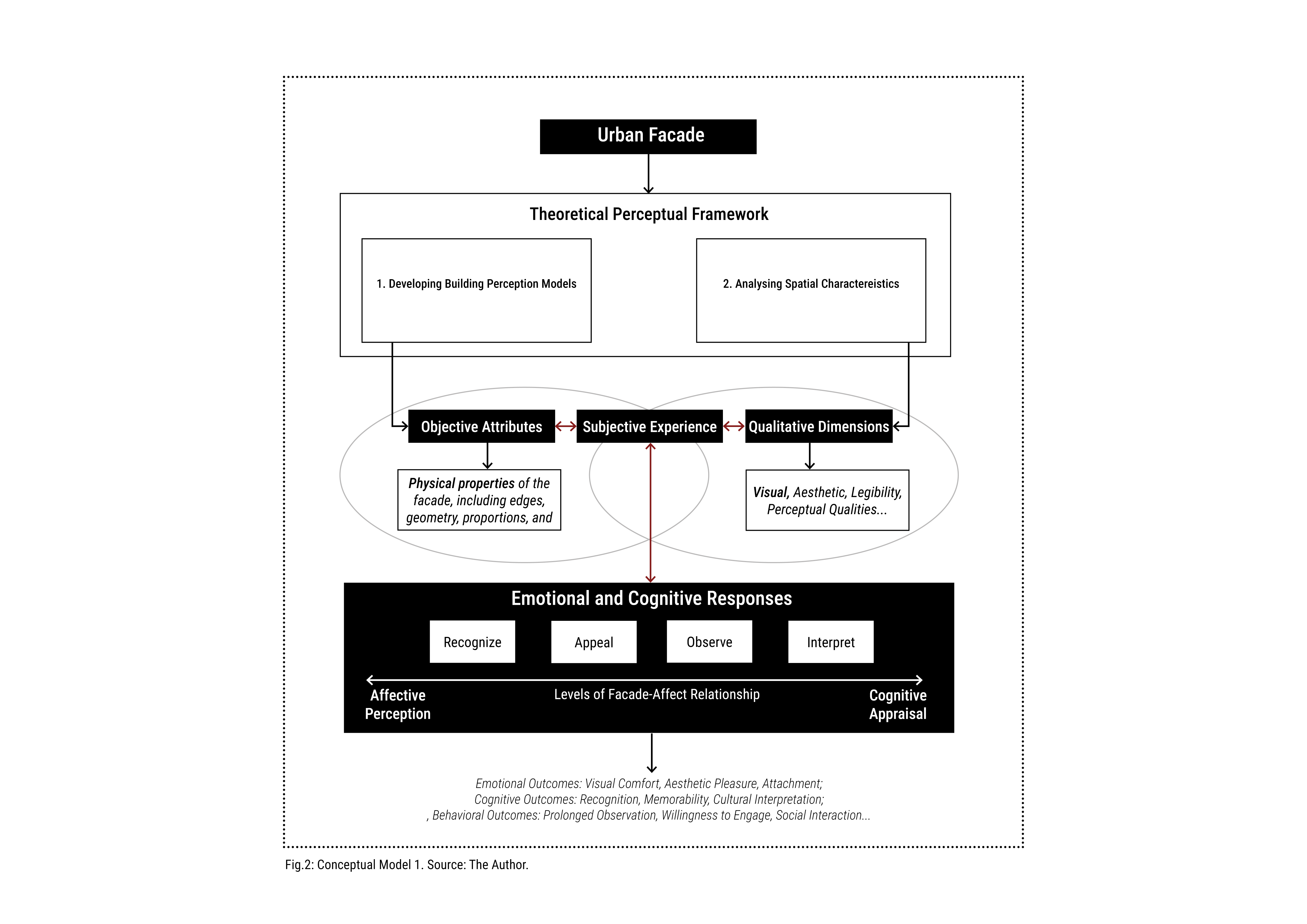}
    \caption{Theoretical perceptual framework for urban facades, illustrating the relationship between objective attributes, subjective experience, qualitative dimensions, and their impact on emotional and cognitive responses. Source: Author.\label{fig7}}
\end{adjustwidth}
\end{figure}

While these properties are valuable as conceptual tools in the critique of facade design, many of their definitions and applications remain imprecise, especially in relation to intangible attributes. For example, mystery, often defined as the arousal of interest or anticipation, depends heavily on subjective interpretation shaped by cultural background and individual experience, which complicates its standardization across studies \citep{Herzog1992, Kaplan1995}. Similarly, enclosure is frequently conflated with physical parameters such as wall height or spacing, obscuring the distinction between its physical and psychological dimensions \citep{Nasar1994, Lindal2013}. Moreover, most frameworks insufficiently account for individual variation and the dynamic, interactive nature of urban environments. In practice, the criteria used to evaluate facade attributes can vary considerably across personal traits, cultural contexts, and situational conditions \citep{Tost2015, Lederbogen2011}. 

Recent advances in empirical research, however, have begun to connect measurable facade features with affective responses. For instance, façade geometry and window-to-wall ratios have been linked to perceived oppressiveness \citep{Wang2024}, adaptive façade systems have been shown to modulate emotional experience through dynamic light and form \citep{Beatini2024}, and perceptual studies of stadium facades highlight relationships between complexity, impressiveness, and preference \citep{Arslan2023}. Further evidence suggests that façade surface properties, such as material roughness and solidity, directly affect stress responses \citep{Christofi2025} while facade color has been shown to influence emotional appraisals and entry decisions \citep{Zhu2024}. Complementary systematic reviews reinforce these findings, underscoring the need for integrative, perception-based approaches to capture affective variations across diverse urban contexts \citep{Shynu2023}.

\subsection{Street-level imagery in human perception measurement} 

Street view imagery (SVI) has transformed the way human perception is studied in urban settings by providing scalable, accurate, and economical tools to analyze building facades. The integration of SVI allows for systematic assessment of facade attributes, linking subjective impressions with objective measurements to evaluate their impact on human perception \citep{Zhang2021, Biljecki2021, Liang2024}. Early studies such as Salesses et al. \citet{Salesses2013} pioneered the use of pairwise comparisons on SVI to examine perceptions of safety, beauty and vibrancy in cities, establishing a foundation for quantitative perceptual analysis. Dubey et al. \citet{Dubey2016} extended this framework with Place Pulse 2.0, a large-scale dataset of six perceptual attributes: depressing, boring, beautiful, safe, lively, and wealthy, allowing systematic mapping of emotional responses and their connection to real-world urban environments. Fig.~\ref{fig7} illustrates an example procedure of such analyses.

Recent improvements in machine learning (ML), particularly Convolutional Neural Networks
(CNNs), have significantly advanced the analytical capabilities of SVI-based perception studies. Fine-tuned models such as ResNet and VGG applied to datasets like Place Pulse 2.0 have enabled the accurate classification of perceptual attributes including safety, beauty, and vibrancy \citep{Dubey2016, Zhang2018}. Generative Adversarial Networks (GANs), especially through the Pix2Pix framework, have further introduced novel pathways for image-to-image translation and synthetic façade generation, expanding the methodological toolkit available for urban perception research \citep{Isola2017}. These advances allow for the extraction and quantification of architectural features such as styles, fenestration patterns, materials, and wall textures, linking them with perceptual variables such as safety or visual interest \citep{Liang2024, Wang2024Oppressiveness, Christofi2025}. Despite these developments, such computational approaches often underrepresent the psycho-physical, cultural, and contextual dimensions that shape localized and subjective urban perception, as highlighted in recent systematic reviews \citep{Shynu2023}.

\subsection{Current literature and studies on perceived facade affect}

\begin{figure}[H]
\begin{adjustwidth}{-\extralength}{0cm} 
    \centering
    \includegraphics[width=\fulllength]{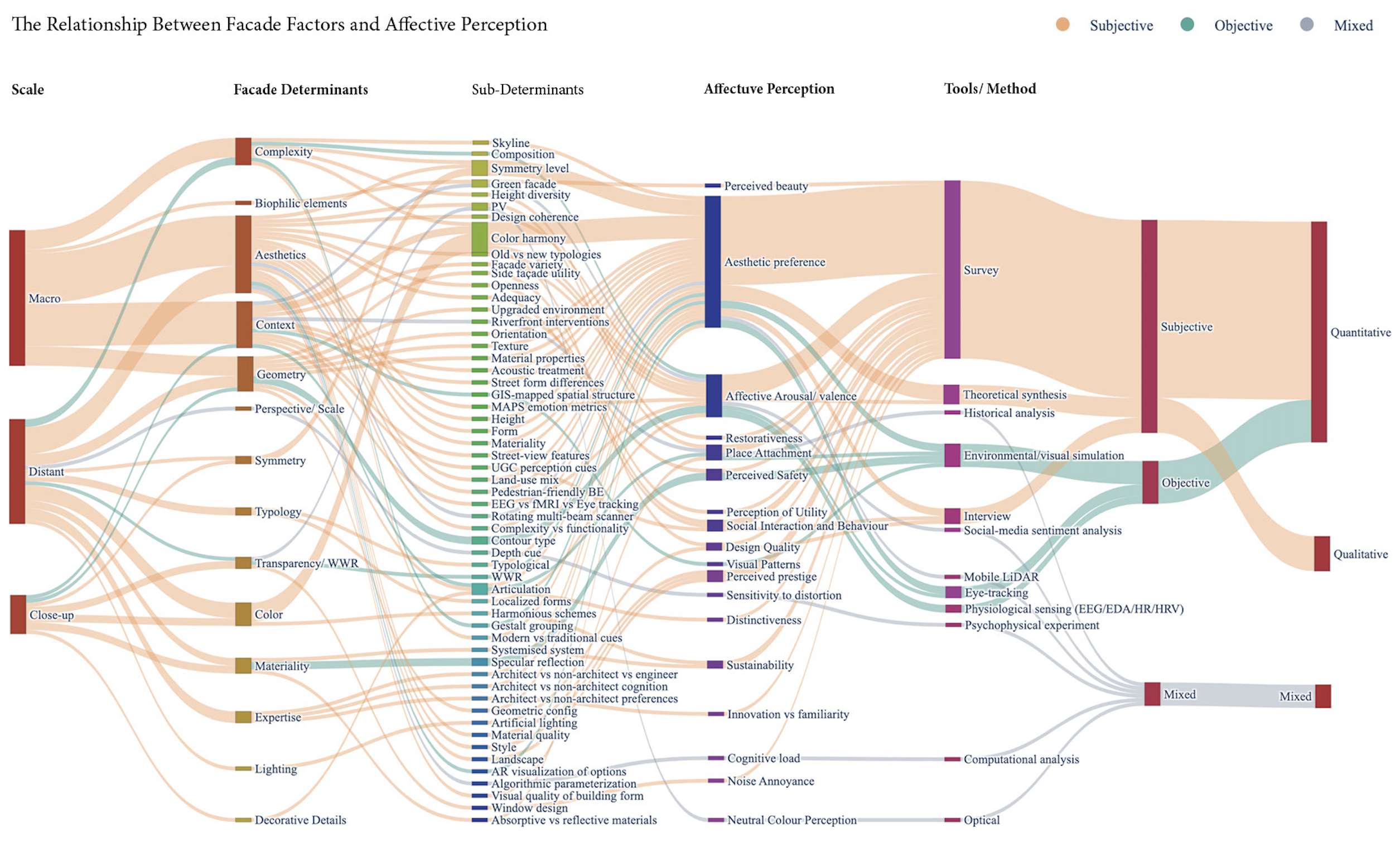}
    \caption{The relationship between facade factors and affective perception. Source: Author.\label{fig8}}
\end{adjustwidth}
\end{figure}

The results of the 61 selected studies are systematically deduced through descriptive and explanatory analysis as part of PRISMA and demonstrate how architectural facade design influences affective perception through a multiscale theoretical framework as ~\ref{fig8}. This multiscale theoretical framework combines macroperspective studies (urban / street scale), distant perspective (building scale), and close-up perspective (detail scale) with the inclusion of affective appraisal processes, including primary and secondary evaluations. By piecing together their findings, a descriptive methodology combining thematic analysis with multiscale views and effective semantic interpretation exposed the link between a facade and human perception.

\subsubsection{Urban/Streetscape Scale (Macro)}
\textit{Perception of the facade within the broader cityscape or urban environment.}

\begin{table}[H]
\tablesize{\scriptsize}
\caption{Studies on affective perception of façades at the urban/streetscape scale (Macro).}
\label{tab:macro}

\begingroup
\renewcommand{\arraystretch}{1.15}
\setlength{\tabcolsep}{5pt}

\newcolumntype{L}{>{\raggedright\arraybackslash}p{0.18\textwidth}}
\newcolumntype{C}{>{\raggedright\arraybackslash}p{0.14\textwidth}}
\newcolumntype{Y}{>{\raggedright\arraybackslash}X}

\begingroup\hyphenpenalty=10000\exhyphenpenalty=10000\sloppy
\begin{adjustwidth}{-\extralength}{0cm}

\begin{tabularx}{\fulllength}{L C C Y Y Y}
\toprule
\textbf{Study} & \textbf{Context} & \textbf{Determinants} & \textbf{Sub-factors} & \textbf{Affective Perception} & \textbf{Tools/Data} \\
\midrule
Gifford et al.\ (2000) \citep{Gifford2000} & City (NA) & Visual complexity & Stylistic cues (modern vs.\ traditional) & Aesthetic judgment (architects vs.\ laypersons) & Questionnaire \\
Heath et al.\ (2000) \citep{Heath2000} & City skyline & Visual complexity & Skyline density; façade composition & Preference; perceived complexity & Survey \\
White \& Gatersleben (2011) \citep{White2011} & Neighborhood (UK) & Biophilic features & Green façades / vegetation & Restorative quality; perceived beauty & Questionnaire \\
Lindal \& Hartig (2013) \citep{Lindal2013} & Neighborhood (EU) & Variation \& scale & Height diversity; façade variety & Restorative perception; preference & Photo-simulation \newline survey \\
Hollander \& Anderson (2020) \citep{HollanderAnderson2020} & City (US) & Urban façade quality & Mixed-use street façades & Affective evaluation (pleasantness, attractiveness) & Structured survey \\
\addlinespace
Abu Qadourah \& Alnusairat (2024) \citep{AbuQadourah2024} & Community (Jordan) & Sustainability aesthetics & PV panel integration on façades & Place attachment; aesthetic evaluation & Structured survey \\
\bottomrule
\end{tabularx}

\end{adjustwidth}
\endgroup
\endgroup
\end{table}

On the macro scale, facades operate as critical components of urban morphology, shaping the visual identity of the cityscape, sociocultural symbolism, and the spatial cognition of pedestrians. Empirical studies demonstrate that façade attributes such as stylistic coherence, skyline complexity, biophilic integration, and height variation significantly influence affective appraisals (Table~\ref{tab:macro}). For example, stylistic cues distinguishing modern from traditional architecture affect aesthetic judgment differently among architects and laypersons \citep{Gifford2000}, while skyline density and façade composition modulate perceived complexity and preference \citep{Heath2000}. The incorporation of green façades and vegetative features enhances restorative quality and perceived beauty \citep{White2011}, and variation in height and scale fosters both restorative perception and preference \citep{Lindal2013}. Recent work has also shown that façade quality along mixed-use urban streets influences cognitive load and affective responses measured through perceptual and behavioral indicators \citep{HollanderAnderson2020}. In sustainable design contexts, the integration of photovoltaic panels into residential façades affects both aesthetic evaluation and place attachment \citep{AbuQadourah2024}. 

More broadly, facades that demonstrate contextual congruence and material coherence strengthen affective resonance, semiotic cohesion, and a sense of place identity \citep{Liang2024, HollanderAnderson2020}. Conversely, incongruous materials or highly reflective surfaces may induce perceptual dissonance, diminishing valence and generating discomfort by fragmenting urban continuity \citep{AbuQadourah2024}. Excessive geometric rigidity or poorly scaled façade compositions can reduce arousal and erode vibrancy in the public realm \citep{Mirabi2020}. Analytical frameworks leveraging geo-spatial intelligence, such as Street View Imagery (SVI) and GIS-based visual-spatial analytics, have become pivotal for quantifying these affective impacts, allowing systematic measurement of façade typologies across diverse urban contexts \citep{Liang2024}.

\subsubsection{Building Scale (Distant)}
\textit{Perception of facade as a single building on the street.}

\begin{table}[H]
\tablesize{\scriptsize}
\caption{Studies on affective perception of façades at the building scale (Distant).} 
\label{tab:building}

\begingroup
\renewcommand{\arraystretch}{1.15}
\setlength{\tabcolsep}{5pt}


\newcolumntype{C}{>{\raggedright\arraybackslash}p{0.14\textwidth}}
\newcolumntype{Y}{>{\raggedright\arraybackslash}X}

\begingroup\hyphenpenalty=10000\exhyphenpenalty=10000\sloppy
\begin{adjustwidth}{-\extralength}{0cm}

\begin{tabularx}{\fulllength}{L C C Y Y Y}
\toprule
\textbf{Study} & \textbf{Context} & \textbf{Determinants} & \textbf{Sub-factors} & \textbf{Affective Perception} & \textbf{Tools/Data} \\
\midrule
Shemesh et al.\ (2017) \citep{Shemesh2017} & Public buildings (Israel) & Geometry & Curvature; convexity; enclosure & Pleasantness; arousal (neural correlates) & EEG \newline questionnaire \\
Aydin \& Mirzaei (2022) \citep{Aydin2022} & University buildings (Iran) & Symmetry & Symmetry index/level & Symmetry preference & Mathematical model \newline survey \\
Ghomeshi \& Jusan (2013) \citep{Ghomeshi2013} & Residential (Iran) & Conceptual design & Typological/semantic attributes & Preference gap (architects vs.\ laypersons) & Questionnaire \\
Yi (2019) \citep{Yi2019} & Office (Korea) & Transparency \& daylight & Window-to-wall ratio; daylight factor & Aesthetic acceptability & Simulation \newline (Energy+, Radiance) \\
\addlinespace
Ilbeigi et al.\ (2019) \citep{Ilbeigi2019} & Residential (Iran) & Proportion/scale & Façade articulation; massing & Cognitive differentiation & Questionnaire \\
Wang et al.\ (2024) \citep{Wang2024} & Churches (China) & Typology & Localized church façade forms & Impression; cognition (affective) & Eye-tracking \\
Zhu et al.\ (2024) \citep{Zhu2024} & Shopping malls (China) & Color & Palette/combination at building scale & Entry decision; approach–avoidance & Structured questionnaire \\
Beder et al.\ (2024) \citep{Beder2024} & Residential (Turkey) & Gestalt principles & Proximity; symmetry; continuity & Aesthetic preference; arousal & Eye-tracking \newline questionnaire \\
Ariannia et al.\ (2024) \citep{Ariannia2024} & Cultural buildings (Iran) & Visual quality & Form semantics; façade quality & Place attachment; affective appraisal & Semantic differential survey \\
\bottomrule
\end{tabularx}

\end{adjustwidth}
\endgroup
\endgroup
\end{table}

At the building scale, facades are regarded as fundamental elements of architectural composition, with particular emphasis on achieving design coherence and visual appeal. Research emphasizes the influence of geometry, symmetry, material articulation, and contextual harmony in enhancing valence by evoking positive affective responses and promoting aesthetic admiration \citep{Shemesh2017, Aydin2022}. Simultaneously, these features contribute to arousal by introducing engaging visual stimuli that capture attention \citep{Aydin2022, Beder2024}. Studies grounded in Gestalt theory reveal that proximity-driven designs generate heightened valence through visual clarity and cognitive efficiency, whereas excessive visual load tends to reduce valence and prolong fixation times \citep{Beder2024}. These findings suggest that symmetry and balanced complexity in facade composition enhance both arousal and valence, fostering perceptions of beauty and emotional engagement.  

Innovative technological tools have significantly advanced the analysis of building-scale facades. Parametric design platforms such as Rhino and Grasshopper, alongside emerging AI tools like Midjourney or Stable Diffusion, enable precise manipulation of facade geometries and material transitions. Semantic differential surveys further capture public affective evaluations, linking facade attributes with emotional dimensions such as arousal and valence \citep{Ariannia2024}. In addition, photogrammetry and eye-tracking methodologies provide granular insights into visual engagement, analyzing fixation patterns and attention distribution on facade elements \citep{Beder2024}. These technologies demonstrate how design attributes are visually processed, highlighting the interplay between stimulating arousal and sustaining positive valence in shaping compelling architectural experiences \citep{Heath2000, Lindal2013}.

\subsubsection{Detail Scale (Close-Up)}
\textit{Perception of facade details visible in close range. }

\begin{table}[H]
\caption{Studies on affective perception of façades at the detail scale (Close-up).}
\label{tab:detail}
\begingroup\hyphenpenalty=10000\exhyphenpenalty=10000\sloppy
\begin{adjustwidth}{-\extralength}{0cm}
\begin{tabularx}{\fulllength}{L L L X X X}
\toprule
\textbf{Study} & \textbf{Context} & \textbf{Determinants} & \textbf{Sub-factors} & \textbf{Affective Perception} & \textbf{Tools/Data} \\
\midrule
Calleri et al. (2017) \citep{Calleri2017} & Public spaces (Italy) & Materiality & Material articulation; surface finish & Affective comfort (auditory–visual) & Acoustic simulation \\
Calleri et al. (2018) \citep{Calleri2018} & Urban blocks (EU) & Materiality \& detailing & Texture; jointing; surface differences & Multisensory aesthetic evaluation & On-site observation \\
Beder et al. (2024) \citep{Beder2024} & Residential (Turkey) & Complexity (Gestalt) & Grouping; detail complexity & Preference; pleasure–arousal & Eye-tracking (pupil, fixation) \\
Vangorp et al. (2013) \citep{Vangorp2013} & Rendering (virtual) & Perspective & Distortion; depth realism & Visual comfort; perceptual realism & Image-based rendering \\
Brzezicki (2013) \citep{Brzezicki2013} & Transparent façades & Transparency & Reflectivity; transmittance & Safety perception; clarity & Visual tasks \\
\addlinespace
Wang, Shen \& Huang (2024)\citep{WangZ2024} & Residential (Shanghai) & Color (detail tones) & Hue/value/chroma at close range & Visual comfort & Eye-tracking (heat map) \\
Montañana et al. (2024) \citep{Montanana2024} & Housing (Spain) & Prestige cues & Detail articulation; ornamentation & Perceived prestige & Structured survey \\
\bottomrule
\end{tabularx}
\end{adjustwidth}
\endgroup
\end{table}

At the detail scale, the emphasis shifts to microlevel elements, including fenestration patterns, articulation, and material treatments, which play a critical role in shaping tactile and visual experiences. Symmetrical and regular fenestration patterns contribute to a sense of harmony and emotional connectivity, effectively improving valence while maintaining a balanced level of arousal \citep{Calleri2017, Calleri2018}. In addition, detailed articulation, such as modular jointing, ornamentation, and textured materiality, amplifies both valence and arousal, fostering dynamic and emotionally engaging architectural spaces \citep{Beder2024, Montanana2024}. Biophilic detailing, for instance the incorporation of vegetation or natural textures into façade systems, further enhances valence by evoking positive affective responses and aesthetic satisfaction, promoting perceptions of beauty and a deeper connection to nature \citep{WangZ2024, White2011}.  

By contrast, facades dominated by reflective or monotonous, featureless surfaces are frequently linked to lower valence and arousal, evoking negative emotional states such as disinterest or discomfort \citep{Brzezicki2013}. The evaluation of such detailed elements has been refined through advanced analytical methodologies. Spectral colorimetric and acoustic simulation tools allow precise assessments of material properties, while virtual reality (VR) and image-based rendering platforms enable immersive exploration of visual and tactile interactions \citep{Vangorp2013}. Complementary qualitative methods, including focus groups and tactile interaction studies, further deepen understanding of how materials, textures, and patterns influence sensory engagement and affective response \citep{Calleri2018}.

\section{Discussion}
\subsection{Key indicators of facade perception (determinants)}

The affective evaluation of facades is mediated not only by their formal determinants but also by the methodological pathways through which these determinants are rendered visible. Complexity, for instance, oscillates between two poles: the measured calm induced by minimalist surfaces and the heightened arousal elicited by ornamented historic facades. Although perceptual fluency theory suggests that moderate complexity facilitates aesthetic preference \citep{Reber2004}, empirical research on skyline articulation reveals that beyond a certain threshold, increased detail becomes a cognitive burden rather than a perceptual affordance \citep{Heath2000, Lindal2013}. Eye-tracking evidence on Gestalt grouping further nuances this by showing that order within complexity, rather than density alone, drives visual preference \citep{Beder2024}. However, these studies remain largely tied to controlled stimuli, which, while valuable, may limit the richness of ecological perception. This tension highlights a methodological asymmetry: surveys and questionnaires capture subjective salience, whereas computational or psychophysiological tools extract quantifiable regularities, often abstracted from lived context.  

A similar disjunction emerges in the study of symmetry and transparency. Mathematical models of symmetry operationalize regularity as a measurable attribute \citep{Aydin2022}, but survey data reveal that laypeople prefer coherence and familiarity over formal precision \citep{Ghomeishi2021}. Transparency studies underscore the same divide: simulations demonstrate the acoustic and daylighting benefits of fenestration \citep{Chau2020, Yi2019}, yet resident reports point instead to comfort and perceived safety as critical outcomes \citep{Brzezicki2013}. In both cases, methodological choice does not simply measure different facets of the same phenomenon; it actively conditions what counts as evidence of affective response.

The determinants of color and materiality illustrate even more strongly the need for integrated approaches. Harmonious chromatic schemes align with sustainability narratives from urban policies \citep{OConnor2006, Oludare2021}, while experimental manipulations of tonal warmth directly shape the behavior of the consumer approach \citep{WangZ2024, Zhu2024}. Acoustic simulations show that porous materials mitigate urban noise \citep{Calleri2018}, but reflective façades – although performing well in energy terms – produce perceptual confusion and safety hazards in user studies \citep{Brzezicki2013}. Meanwhile, vegetated façades are consistently rated as restorative and beautiful \citep{White2011}, highlighting how ecological integration activates affective pathways that cannot be captured by performance metrics alone.

\subsection{Application of urban facades evaluation (Methods)}

The evaluation of urban façades has evolved significantly, with advanced methodologies increasingly replacing traditional observational tools to better quantify their affective perception. Early approaches, such as field surveys, interviews, and questionnaires, provided valuable qualitative insights into how people perceived façades in terms of aesthetics, coherence, and contextual relevance. However, these approaches were restricted by their reliance on subjective self-reports, which are often susceptible to biases and limited scalability \citep{Nasar1994, Dunstan2005, Shynu2023}. 

Recent advances have seen the introduction of computational tools, including photogrammetry, parametric modeling software (e.g., Rhino and Grasshopper), and lighting simulations, which enable façade characteristics to be analyzed with precision \citep{Calleri2017, Beder2024, DeLaColina2016, Ghozatlou2024}. By facilitating iterative testing in controlled environments, these tools generate quantitative insight into how specific design features influence human perception. From the perspective of material analysis, parametric modeling and daylight simulations have been effectively used to assess reflectivity and light absorption, producing actionable data to optimize façade designs in practice \citep{Yi2019, Chau2020}. Midjourney and Stable Diffusion are also increasingly applied for controlled visual manipulations \citep{Valentine2025}. However, when viewed in the context of urban environments, the controlled nature of these methods often isolates them from the dynamic variability of real-world settings, thus constraining their ecological validity \citep{Gregorians2022, Shynu2023}. A further step forward is marked by virtual reality (VR) and eye-tracking technologies, combined with Agent-Based Modeling (ABM), which allow researchers to explore visual attention and affective responses within simulated urban scenarios \citep{Shemesh2017, Zhang2021}. Although these tools facilitate the tracking of visual focus and the measurement of affective reactions to design elements such as color, transparency, and texture \citep{WangZ2024, Brzezicki2013}, the lack of demographic diversity among participants continues to limit the generalizability of findings \citep{Zhu2024, Aydin2022}.

Understanding has been further deepened by qualitative frameworks based on Gestalt principles, which investigate how individuals cognitively and emotionally process façade elements. For example, mental effort and user satisfaction have been shown to be influenced by design simplicity or complexity, as revealed by eye-tracking and EEG-based analyses of perceptual load \citep{Shemesh2017, Beder2024}. However, these contributions often originate from laboratory experiments, in which real-world variables such as ambient noise, thermal comfort, and pedestrian flow are excluded. From a practical design perspective, this omission underscores the difficulty of translating controlled experimental results into applicable insights for dynamic urban contexts, highlighting an ongoing challenge for the field.

\section{Limitation}
\subsection{Gaps in Analytical Evaluation for Urban Facades}
A significant limitation in the analysis of urban facades is the absence of comprehensive and systematic frameworks capable of capturing the nuanced relationship between facade attributes and their impact on human perception and affective responses.  Current methodologies often treat facades as homogeneous entities, in which they fail to disaggregated their constituent elements or analyze their affective interactions with precision. For example, while Hollander and Anderson (2020) emphasize the emotional appeal of 'high-quality' facades, their analysis does not delineate the specific attributes that contribute to these perceptions, resulting in generalized conclusions with limited empirical applicability. This lack of specificity has led to fragmented and inconsistent findings, thereby impeding the development of robust theoretical and practical frameworks. 

In addition, the dichotomy between tangible and intangible dimensions restricts the field's ability to account for the full spectrum of perceptual and emotional responses elicited by facades. Intangible attributes, usually described as conceptual or visual features of facades, such as coherence, mystery, and cultural symbolism, remain significantly underexplored relative to tangible or morphological features, such as symmetry or materiality. Although, for example, symmetry is often associated with coherence, studies rarely contextualize this association. Future research should prioritize integrative methodologies that link quantifiable attributes, such as color and texture, with higher-order perceptual constructs, embedding these analyses within cultural and urban contexts to generate a cohesive and predictive understanding of facade aesthetics and their affective impact.

\subsection{Limited Integration of Mixed-Method Approaches}

What emerges across these strands is less a set of isolated determinants than a triangular relationship: facade attributes, affective outcomes, and methodological tools are co-constructed. Surveys and interviews tend to foreground subjective well-being and attachment; simulations and modeling quantify environmental performance; physiological and behavioral tools, such as eye-tracking or EEG, reveal implicit cognitive load. Crucially, these modes of measurement often generate divergent, sometimes conflicting accounts of the same facade feature. The task of future research, therefore, is not merely to accumulate evidence across scales but to reconcile these epistemic layers, developing integrative models in which facade complexity, transparency, color, and materiality are understood not simply as physical determinants but as affective mediators whose impact is contingent on the methodological lens through which they are observed.

\section{Conclusion}
This review addressed the research objectives by systematically synthesizing evidence on how architectural facades influence human affective perception, particularly through arousal and valence, using a PRISMA-guided methodology. The findings demonstrate that facade attributes, including contextual harmony and luminance gradients at the urban scale, symmetry and historical complexity at the building scale, and intricate textures and bibliophilic elements at the detail scale, are critical determinants of affective responses. These attributes distinctly shape emotional engagement and aesthetic evaluation across spatial scales. In addition, interdisciplinary methodologies that combine computational tools with empirical studies effectively connect these attributes to measurable affective outcomes, providing clarity on how facade design mediates emotional engagement and supports psychological well-being.

These insights highlight the fundamental role of facade design in improving urban livability, well-being, and aesthetic appeal, offering valuable implications for evidence-based design strategies aimed at creating affectively supportive urban environments. However, the findings should be interpreted considering limitations such as fragmented integration of qualitative and quantitative approaches, inadequate exploration of intangible facade attributes such as coherence and mystery, and underrepresented contextual and cultural dimensions in existing studies. Future research should address these gaps by developing comprehensive frameworks that link subjective affective appraisals with objective facade metrics and by exploring culturally diverse and context-sensitive methodologies to further enrich our understanding of facade-affect dynamics.

\vspace{6pt} 

\authorcontributions{
Conceptualization, Chenxi Wang and Michal Gath-Morad; 
methodology, Chenxi Wang; 
software, Chenxi Wang; 
validation, Chenxi Wang and Michal Gath-Morad; 
formal analysis, Chenxi Wang; 
investigation, Chenxi Wang; 
resources, Chenxi Wang; 
data curation, Chenxi Wang; 
writing-original draft preparation, Chenxi Wang; 
writing-review and editing, Chenxi Wang and Michal Gath-Morad; 
visualization, Chenxi Wang; 
supervision, Michal Gath-Morad; 
project administration, Michal Gath-Morad; 
funding acquisition, Michal Gath-Morad. 
All authors have read and agreed to the published version of the manuscript.
}

\funding{ ``This research received no external funding''}

\institutionalreview{The study was conducted in accordance with the Declaration of Helsinki, and approved by the Ethics Committee of the University of Cambridge, Department of Architecture (protocol code AHA/PG/ST24100021, approved in 2024).}

\informedconsent{Not applicable.}

\dataavailability{No new data were created in this study. The literature dataset supporting the findings of this review is available within the article and its Supplementary Materials.}

\acknowledgments{The author would like to thank Dr. Michal Gath-Morad for her intellectual support and constructive feedback, which greatly contributed to the refinement of this manuscript.}

\conflictsofinterest{The author declares no conflict of interest.}

\abbreviations{Abbreviations}{
The following abbreviations are used in this manuscript:
\\

\noindent 
\begin{tabular}{@{}ll}
CNN & Convolutional Neural Network\\
GANs & Generative Adversarial Networks\\
PRISMA & Preferred Reporting Items for Systematic Reviews and Meta-Analyses\\
SVI & Street View Imagery\\
UGS & Urban Green Space
\end{tabular}
}

\reftitle{References}

\PublishersNote{}
                
\bibliography{reference}                      
\end{document}